# Universal enhancement of vacancy diffusion by Mn inducing anomalous Friedel oscillation in concentrated solid-solution alloys

Huaqing Guan[a], Shaosong Huang[a*], Fuyang Tian[b], Chenyang Lu[c*], Qiu Xu[d], and Jijun Zhao[a]

[a] Key Laboratory of Materials Modification by Laser, Ion and Electron Beams (Dalian University of Technology), Ministry of Education, Dalian, 116024, China
[b] Institute for Applied Physics, University of Science and Technology Beijing, Beijing, 10083, China
[c] Department of Nuclear Science and Technology, Xi'an Jiaotong University, Xi'an, 710049, China
[d] Institute for Integrated Radiation and Nuclear Science, Kyoto University, Osaka, 5900494, Japan

## Abstract

We present a proof-of-principle demonstration of a universal law for the element Mn, which greatly enhances vacancy diffusion through an anomalous Friedel Oscillation effect in a series of Ni-based concentrated solid-solution alloys, regardless of the type of atom involved. The antiferromagnetic element Mn possesses a unique half-filled *3d* electron structure, creating split virtual bound states near the Fermi energy level and producing a large local magnetic moment after vacancy formation. The resultant electron spin oscillations reduce the number of electrons involved in charge density oscillations, destroying charge screening and lowering potential interaction at the saddle point between the vacancy and diffusing atom. This ultimately facilitates vacancy diffusion by reducing energy level variations of conduction band electrons during the diffusion process. These findings offer valuable insights into atom diffusion mechanisms and open up new avenues for manipulating defect properties through unique element design, thereby enabling the creation of high-performance alloys in a broad range of fields.

In the field of materials science, uncovering the intricate relationship between the structure and the physical properties, as well as discovering new laws, is the key to unlocking a deeper understanding of the subject. Since the Bronze Age, metal alloying has been used as a strategy to improve material properties [1]. Recent advancements in this field include Concentrated Solid-Solution Alloys (CSAs) and High Entropy Alloys (HEAs) [2-4], which offer exceptional properties that surpass those of traditional alloys. The high concentrations of multiple elements in these alloys provide a vast composition space, but also present a challenge in discovering the underlying physical mechanisms that govern their properties. The concept of "sluggish" diffusion [5], a core effect of HEAs, is hotly debated [6,7], with recent studies [8,9] suggesting that the type of elements in the alloy plays a greater role than configurational entropy. Additionally, the specific elements chosen in a CSA significantly affect the alloy's response to irradiation, hydrogen embrittlement, and other properties [10-12]. The movement of atoms in crystalline solids is often facilitated by vacancy point defects, making them a crucial factor in the diffusion process. As a result, the rate of microstructural evolution and phase transformations in CSAs is greatly influenced by the diffusivity of these vacancies. However, the complex local atomic environments make it difficult to understand the effect of the element-specific chemical complexity on point defect evolution in CSAs. Conventional methods such as mean-field averaging [13] are not suitable for this task, and existing microscopies lack the capability for element-resolved or time-resolved kinetics. Improving our understanding of the relationship between the presence of special solute atoms and point defect physics is critical to advance the development of high-performance alloys with targeted property design. In this letter, we successfully demonstrated the significant impact of specific solute atoms on vacancy diffusion from a metal electron theoretical perspective, through first-principal calculations in ternary alloys $Ni_{0.6}Co_{0.2}Co_{0.2}$, $Ni_{0.6}Co_{0.2}Fe_{0.2}$, and $Ni_{0.6}Co_{0.2}Mn_{0.2}$ CSAs. To address the limitations of studying unique site-to-site atomic environments, we utilized a Density Functional Theory (DFT) approach based on the Similar Atomic Environment (SAE) method [14-17]. This technique transforms structural modeling into a minimization problem by optimizing the configuration, allowing for a more precise characterization of short-range order within solid solutions [17]. Consequently, it can provide a more comprehensive understanding of the relationship between chemical electronic structures and defect evolution, enabling a deeper insight into the properties of solid solutions.

The methodology used is thoroughly explained in the supplementary materials. Specifically, the Climbing-Image Nudged Elastic Band (CI-NEB) method [18] is used to determine the vacancy migration energy. We remove one atom from the structure and calculate the migration barrier for all neighboring atoms jumping into the created vacancy. The supercell of the three alloys, namely $Ni_{0.6}Co_{0.2}X_{0.2}$ (where X= Co, Fe, Mn, respectively), were generated with the same size and atomic environments except for the X solute atoms. To isolate the influence of the local X element on vacancy diffusion, vacancies were randomly selected at the same position in all alloys, as shown in **Fig. S1**. **Fig. S2** illustrates the vacancy diffusion energies, categorized by chemical species (i.e., Ni, Co, Fe, or Mn). **Fig. 1** shows one of the cases of vacancy diffusion as a representative. The results are significant, as all elements in the $Ni_{0.6}Co_{0.2}Mn_{0.2}$ alloy show notably lower diffusion energies, averaging 0.4 eV less compared to those in the $Ni_{0.6}Co_{0.2}Co_{0.2}$ and $Ni_{0.6}Co_{0.2}Fe_{0.2}$ alloys. These findings suggest that the effect of Mn on vacancy diffusion differs significantly from those of Co and Fe. The diffusion energy, calculated using the CI-NEB method, represents the energy barrier that must be overcome to move from the initial point (IP) to the saddle point (SP) due to the potential difference between the two points [18]. **Fig. 2** presents the potential energy surface (PES) of identical diffusion path in the alloys by using the same reference standard. It is evident that the potential energy at the SP position in NiCoMn is markedly lower than that in the other two alloys. To comprehend this particular result, we have established a correlation between the electronic structure properties and the PES, and we will now elaborate on this connection in the following discussions.

The diffusion process is determined by the interaction between the diffusing atom and its first nearest neighbor (1nn) atoms, as well as the interaction between the diffusing atom and the vacancy (I-V interaction). In this study, the d-electrons were found to be highly localized and chemically inactive due to the closed d-shell. Even at the SP, which is the closest position to the 1nn atoms in the diffusion process, the diffusing atom experiences minimal interactions with the 1nn atoms, as evidenced by **Fig. S3** in the supplementary materials. This means the I-V interaction plays a decisive role in the diffusion process, and it can be expressed using the following equation:

$$V_{I-V}(r) = h\phi_{I-V}(r) = \frac{1}{2}h(V_I(r) + V_V(r)) \tag{1}$$

where h is the concentration-dependent factor, $V_I(r)$ and $V_V(r)$ is the potential induced by the atom and vacancy, respectively. Considering that the diffusing atoms are all Ni in the three alloys, the $V_I(r)$ is the same. Thus, $V_{I-V}(r)$ is mainly correlated with the $V_V(r)$, which can be calculated from the charge change $\triangle\rho$ after the vacancy formation:

$$V_V(r) = \int \frac{\Delta\rho(r)}{|r-r_0|} dr_0 \tag{2}$$

Generally, the $\triangle\rho$ exhibits an oscillatory nature with increasing distance from the defect, known as Friedel Oscillation (FO) [19,20], which was also recently observed experimentally on an iron surface [21]. According to the FO theory, at a distance r from the vacancy, $\triangle\rho$ has the standard form [22]

$$\Delta\rho(r) = A\frac{\cos(2k_F r + \Phi)}{r^3} \tag{3}$$

Where both the amplitude A and the phase $\Phi$ are dependent on the vacancy potential and on the host metal. **Fig. 3 (Left)** present a comparison between the actual changes in electron charge density (red line) following vacancy formation and the fitting results (black line) along <001> directions. Three-dimensional visualizations of the deformation charge density, with the same isosurface value of 0.08 e/Å are also provided in **Fig. 3 (Right)**. The results show that in NiCoCo and NiCoFe alloys, vacancy formation leads to a typical FO that is limited to the 1nn atoms surrounding the vacancy, as can be seen in the cases of many different types of elements in pure Al [23]. In contrast, the screening electric charge in NiCoMn is destroyed, as depicted in the bottom left of **Fig. 3**, the electron oscillations can extend to more than 4nn, with no directional variability, indicating a much different $V_V(r)$.

The electrons involved in the $V_V(r)$ are frequently confined in the first Brillouin region of the vacancy in the inverted lattice, which coincides with the Wigner-Seitz (WS) cell [24]. On the other hand, the SP is situated at the boundary of the WS cell, which means that the potential energy at the SP can be attributed to the effect of the occupied electrons within the WS cell. The Density of States (DOS) can be calculated as a measure of the number of electrons occupying the energy states within the generated vacancy [25],

$$\Delta\rho = \Delta N(E) = N_P(E) - V_V(E) \tag{4}$$

Where $N_P(E)$ and $N_V(E)$ are the integrated DOS with and without vacancy, respectively. Using **Eq. (4)**, the changes in local integrated DOS below the Fermi level after Ni-vacancy formation in the CSAs are investigated, as shown in **Table 1**. According to the Friedel sum-rule, the value of Δρ should be close to the excess charge generated by the atom producing the vacancy [19]. In our case, the excess charge for Ni ($3d^84s^2$) is 10. Consistent with the rule, the value of Δρ was found to be approximately 10 in NiCoCo and NiCoFe alloys. However, in NiCoMn, the value of Δρ was comparatively lower, measuring at 6.71. The decrease in Δρ in NiCoMn suggests that the electrons within the WS cell are reduced, which directly leads to a lower perturbation potential at the SP position, making it easier for the vacancy to diffuse.

To gain insight into the physical mechanisms behind the observed decrease in $V_V(r)$ in NiCoMn, we conducted an examination of the local partial density of states (PDOS) of the 1nn elements surrounding vacancies in the three alloys, as illustrated in **Fig. 4**. In the case of Co and Fe, the majority-spin *3d* band centers are aligned with those in Ni and are located far from the Fermi surface. This phenomenon is the result of minimizing the kinetic energy. On the other hand, for Mn, the PDOS band center shifts towards the Fermi energy level due to the unique presence of single-occupied *3d* electrons that are required for maintaining charge neutrality [26]. As additional single electrons occupy the Fermi level, a highly hybridized interaction between the local moments and electrons occurs, leading to the injection of new electronic states into the conduction sea. This behavior is consistent with the magnetic 'RKKY' interaction [27,28], which produces an apparent splitting of the virtual bound state (VBS) near the Fermi energy level. The concept of VBS was initially introduced by Friedel [19] to address energy level mismatch in dilute alloys. Its existence was verified by C. Berthier and M. Minier through electric field gradient experiments [29,30]. **Fig. 4** displays the results of our analysis, which confirm the presence of split VBS near the Fermi surface in the Mn PDOS after the formation of vacancies. This observation provides evidence of the RKKY's role in the behavior of electronic states, which is consistent with the unique strong interaction between vacancy and 1nn Mn atom observed by L. Messina [31]. After the vacancy formation, we observed a substantial increase in the total magnetic moment of the three 1nn Mn atoms around the vacancy from -1.17 μ$_B$ to -2.38 μ$_B$. Research conducted by D. Billington and S. Mu et al. [26] also found that Mn exhibits a disordered local moment state with large magnetic moments aligned parallel and antiparallel. The introduction of a large local magnetic moment in the Mn atoms causes electronic spin density oscillations in addition to electron density oscillations. Consequently, electrons near the Fermi energy level become confined in VBS. Notably, we found that the change in the integral DOS of the 1nn Mn atoms within the VBS width, as depicted in the small plot in **Fig. 4**, was approximately 3.1 after the vacancy formation. This value aligns with the number of electrons that are not involved in the charge density oscillation due to the vacancy, which is 3.29 (10 minus 6.71). This finding provides further confirmation of the previous discussions.

In summary, the logical physical process can be explained as follows: due to the distinctive electronic structure of Mn, some electrons are limited in their participation in the charge selective oscillations of RKKY. This limitation reduces the number of vacancies involved in the charge density oscillations, resulting in a weaker effect of vacancies on the potential energy at the SP position. As a result, the potential energy difference between the IP and the SP of the diffusing atom decreases, which ultimately leads to an acceleration of the diffusion of vacancy.

As density functional theory (DFT) calculations are typically performed at 0 K, it's important to understand how temperature affects vacancy diffusion behavior. The diffusion activation energy (Q) is the sum of the diffusion energy barrier ($E_b$) and the vacancy formation energy adjacent to a solute ($E_f$). For Ni-vacancies in NiCoCo, NiCoFe, and NiCoMn alloys, $E_f$ is nearly the same as 1.61 eV, 1.60 eV, and 1.58 eV, respectively. This means that the diffusion activation energy Q is primarily determined by the diffusion energy barrier. Previous work by Nowick and Dienes [32] showed that the temperature dependence of the enthalpy for diffusion is at most $K_BT$. Despite reaching temperatures as high as 800 K, the thermal energy represented by $K_BT$ remains relatively low, around 0.06 eV. This is significantly smaller than the average diffusion barrier required for vacancy movement from a hollow to a bridge site in NiCoMn, by a factor of nearly 15. This suggests that the difference in vacancy diffusion energies among the three alloys is to some extent independent of temperature. Although elevated temperatures cause spins disorder and weaken the overall magnetic moments, it's important to note that local magnetic moments still persist for individual defects. This is due to the Coulomb interactions that contribute to the formation of durable local magnetic moments, which are independent of temperature. Even at high temperatures, Mn can form a spin-glass state [26,33], indicating that the Mn atoms still maintain strong electron interactions, including magnetic exchange interaction. As a result, the local magnetic interactions continue to have an impact on vacancy diffusion barriers, even at elevated temperatures.

Traces of Mn's influence on vacancies can be found in various studies [34-38]. For instance, in NiCoFeCrMn, the DFT calculated average vacancy diffusion barrier is about 0.82 eV [34], which is significantly lower than those in NiCoCr and NiCoFeCr [35]. Monte Carlo simulations show that the diffusion barriers of vacancies in NiCoFeCrMn HEAs are concentrated around 0.8 eV, and lower than those of the counterpart elements in the bulk [36,37]. Mn-containing alloys also exhibit significantly accelerated macro-level diffusion of vacancies. In $Ni_x(CoCrFeMn)_{100-x}$ alloys, higher concentrations of Mn lead to lower diffusion barriers of vacancies [37]. Tracer diffusion experiments conducted by Dabrowa et al. [38] reveal that Mn-containing HEAs, such as CoFeMnNi and CoCrMnNi, exhibit faster diffusion kinetics than most other FCC matrices. While it was suggested that Mn could lower the formation energy of vacancies and increase their equilibrium concentration, our previous work [15] contradicts this theory. In fact, we found that Mn-containing alloys exhibit a higher vacancy formation energy than those without Mn. Instead, the Mn-induced anomalous Friedel oscillation is likely the main cause of accelerated vacancy diffusion, as discussed above.

In all, we present a proof-of-principle mechanism that demonstrates a correlation between the dynamic changes of point defects induced by different elements and the intrinsic macro-property behavior. Specifically, in the case of the half-filled *3d* electronic structure Mn, can lead to the destruction of the electron screening effect by an "anomalous FO", reduce the variation of the conduction band electron energy levels during the diffusion process, and ultimately accelerate the diffusion of vacancies. The research highlights the potential for developing advanced materials with desirable performance by tuning defect properties, within a broad range of metallic systems.

This research was supported by the National Natural Science Foundation of China under the number of 12275045, 51771015, and 12075179.

# References

[1] R. E. Hummel, Springer, 2004.
[2] D. B. Miracle, O. N. Senkov, Acta Mater. 122, 448 (2017).
[3] B. Cantor, I. T. H. Chang, P. Knight, A. J. B. Vincent, Mater. Sci. Eng. A. 375, 213 (2004).
[4] M. H. Tsai, J. -W. Yeh, Mater. Res. Lett. 2, 107 (2014).
[5] K. Y. Tsai, M. H. Tsai, J. -W. Yeh, Acta Mater. 61, 13 (2013).
[6] A. Paul, Scripta Mater. 135, 153 (2017).
[7] M. Vaidya, K. G. Pradeep, B. S. Murty, G. Wilde, S. V. Divinski, Acta Mater. 146, 211 (2018).
[8] M. Vaidya, S. Trubel, B. S. Murty, G. Wilde, S. V. Divinski, J. Alloys Compd. 688, 994 (2016).
[9] E. J. Pickering, N. G. Jones, Int. Mater. Rev. 61, 3 (2016).
[10] Y. Zhang, G. M. Stocks, K. Jin, C. Lu, H. Bei, B. C. Sales, L. Wang, L. K. Beland, R. E. Stoller, G. D. Samolyuk, M. Caro, A. Caro, and W. J. Weber, Nat. Commun. 6, 8736 (2015).
[11] Y. Zhang, Y. N. Osetsky, W. J. Weber, Chem. Rev. 122, 789 (2022).
[12] K. Jin, C. Zhang, F. Zhang, H. Bei, Mater. Res. Lett. 6, 5 (2018).
[13] J. W. Negele, Rev. Mod. Phys. 54, 4 (1982).
[14] F. Tian, D. Y. Lin, X. Gao, Y. F. Zhao, and H. F. Song, J. Chem. Physical. 153, 034101 (2020).
[15] H. Guan, S. Huang, J. Ding, F. Tian, Q. Xu, and J. Zhao, Acta Mater. 187, 122 (2020).
[16] G. Kresse, J. Hafner, Phys. Rev. B 49, 14251 (1994).
[17] G. Kresse, J. Furthmüller, Comput. Mater. Sci. 6, 15 (1996).
[18] G. Henkelman, B. P. Uberuaga, H. Josson, J. Chem. Phys. 113, 22 (2000).
[19] J. Friedel, II Nuovo Cimento 7, 2 (1958).
[20] J. Friedel, The London, Edinburgh, and Dublin Philosophical Magazine and Journal of Science 43, 153 (1952).
[21] T. Mitsui, S. Sakai, S. Li, T. Ueno, T. Watanuki, Y. Kobayashi, R. Masuda, M. Seto, H. Akai, Phys. Rev. Lett. 125, 236806 (2020).
[22] J. S. Galsin, Springer Science & Business Media, (2002).
[23] P. A. Sterne, J. van Ek, R. H. Howell, Compu. Mater. Sci. 10, 306 (1998).

[24] A. Kokalj, J. Mol. Graphics Modell. 17, 3 (1999).

[25] J. S. Galsin, Springer Science & Business Media, (2002).

[26] S. Mu, G. D. Samolyuk, S. Wimmer, M. C. Troparevsky, S. N. Khan, S. Mankovsky, H. Ebert, and G. M. Stocks, npj Comp. Mater. 5, 1 (2019).

[27] M. A. Ruderman, C. Kittel, Phys. Rev. 96, 99 (1954).

[28] S. S. P. Parkin, D. Mauri, Phys. Rev. B 44, 7131 (1991).

[29] C.Berthier, M. Minier, J. of Phys. F: Metal Physics 3, 1268 (1973).

[30] C. Berthier, M. Minier, J. of Phys. F: Metal Physics 7, 515 (1977).

[31] L. Messina, M. Nastar, N. Sandberg, P. Olsson, Phys. Rev. B 93, 184302 (2016).

[32] A. S. Nowick, G. J. Dienes, Phys. Status Solidi 24, 461 (1967).

[33] M. Ali, P. Adie, C. H. Marrows, D. Greig, B. J. Hickey, R. L. Stamps, Nat. Mater. 6, 70 (2005).

[34] S. L. Thomas, S. Patala, Acta Mater. 196,144 (2020).

[35] S. Zhao, T. Egami, G. M. Stocks, Y. Zhang, Phys. Rev. Mater. 2, 013602 (2018).

[36] T. Yang, C. Li, S. J. Zinkle, S. Zhao, H. Bei, Y. Zhang, J. Mater. Res. 33, 19 (2018).

[37] J. Kattke, D. Utt, M. L. Brocq, A. Fareed, D. Gaertner, L. Perriere, L. Rogal, A. Stukowski, K. Albe, S. V. Divinski, G. Wilde, Acta Mater. 194, 236 (2020).

[38] J. Dabrowa, M. Zajusz, W. Kucza, G. Cieslak, K. Berent, T. Czeppe, T. Kulik, M. Danielewski, J. Alloys Compd. 783, 30 (2018).

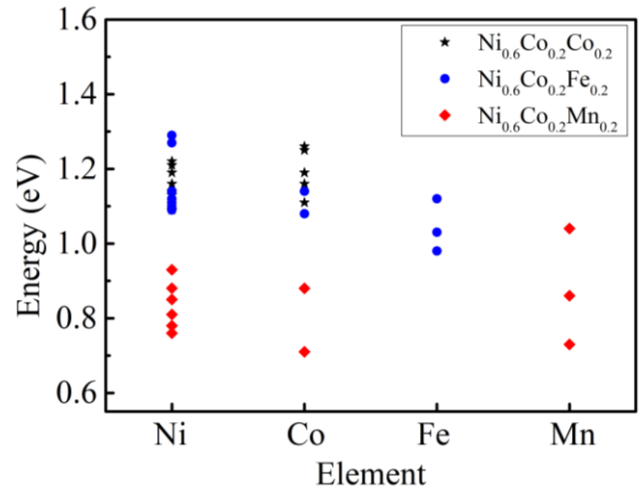

**Fig.1.** The diffusion energy barrier of 1nn atoms exchanging with vacancy in the Ni$_{0.6}$Co$_{0.2}$Co$_{0.2}$, Ni$_{0.6}$Co$_{0.2}$Fe$_{0.2}$, and Ni$_{0.6}$Co$_{0.2}$Mn$_{0.2}$ alloys.

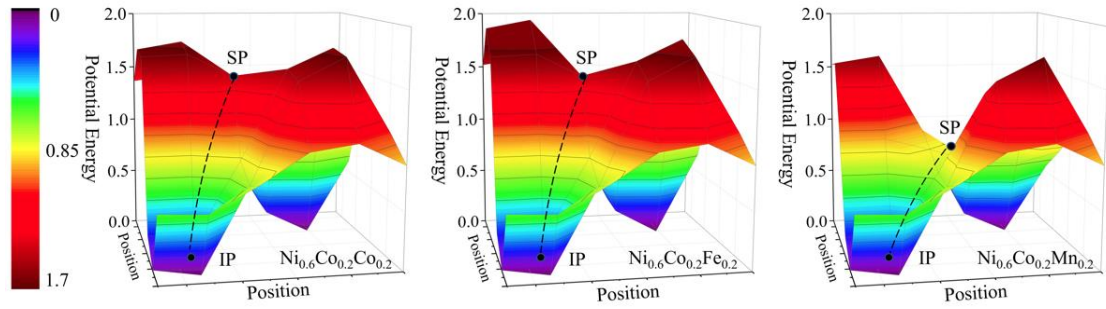

**Fig.2.** The potential energy surface (PES) of identical diffusion path in the Ni$_{0.6}$Co$_{0.2}$Co$_{0.2}$, Ni$_{0.6}$Co$_{0.2}$Fe$_{0.2}$, and Ni$_{0.6}$Co$_{0.2}$Mn$_{0.2}$ alloys by using the same reference standard. Atom diffuses from the initial point (IP) along the dotted line to the saddle point (SP).

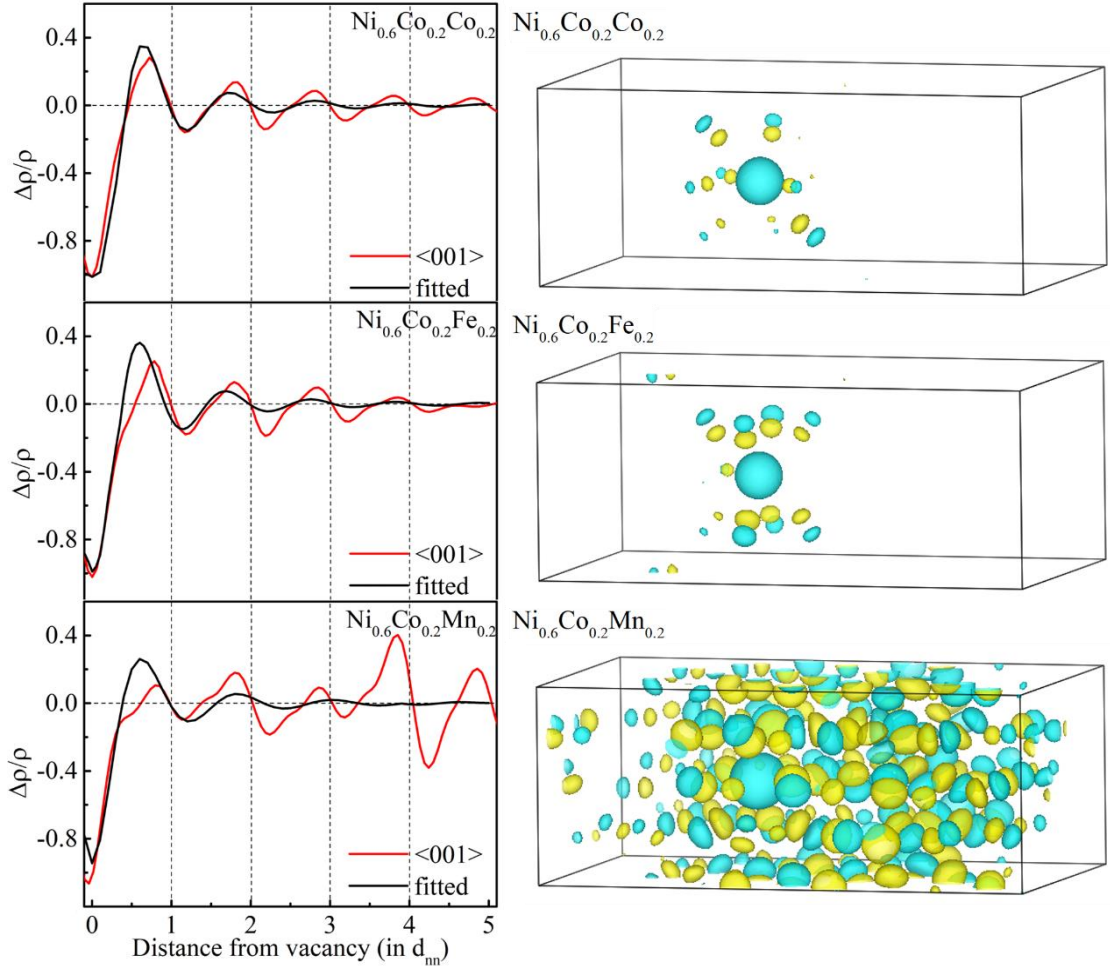

**Fig.3. (Left):** The charge density difference before and after vacancy formation along <001> directions for the same position generated vacancy in $Ni_{0.6}Co_{0.2}Co_{0.2}$, $Ni_{0.6}Co_{0.2}Fe_{0.2}$, and $Ni_{0.6}Co_{0.2}Mn_{0.2}$ supercell, respectively. Distance from the vacancy is in scale of the nearest-neighbor atom-atom distance $d_{nn}$=2.45 Å in three alloys. The actual (red line) and the fitted result (black line) are presented. **(Right):** The calculated three-dimensional deformation charge density with the generation of a vacancy in the alloys with the same isosurface of 0.08 e/Å.

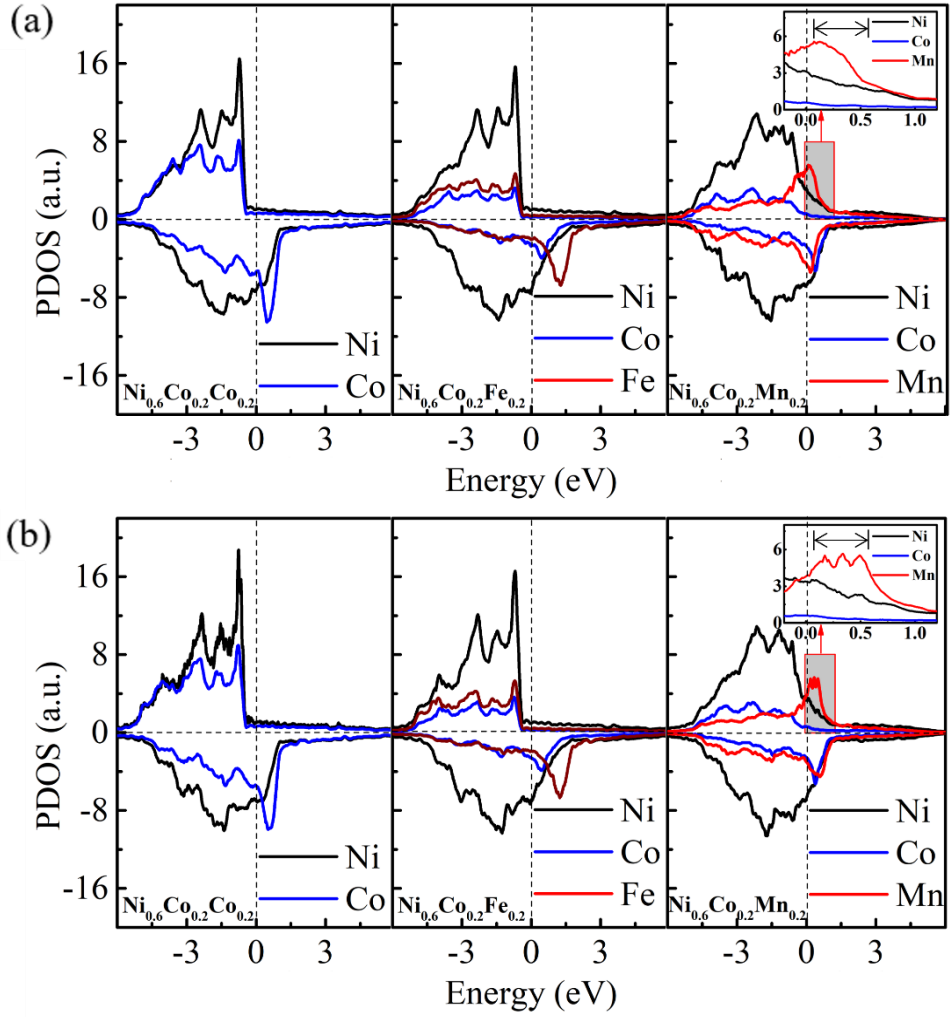

**Fig.4.** The PDOS of the 1nn atoms around the vacancy in $Ni_{0.6}Co_{0.2}Co_{0.2}$, $Ni_{0.6}Co_{0.2}Fe_{0.2}$, and $Ni_{0.6}Co_{0.2}Mn_{0.2}$ alloys. **(a)** and **(b)** are the PDOS of 1nn atoms before and after vacancy formation, respectively.

**Table.1.** The integral density of states (DOS) of vacancies and their first-neighboring atoms in $Ni_{0.6}Co_{0.2}Co_{0.2}$, $Ni_{0.6}Co_{0.2}Fe_{0.2}$, and $Ni_{0.6}Co_{0.2}Mn_{0.2}$ alloys, both before and after the vacancy formation process.

|  | Number of electron charge | | |
| --- | --- | --- | --- |
|  | $N_I(E)$ | $N_0(E)$ | △ |
| $Ni_{0.6}Co_{0.2}Co_{0.2}$ | 115.96 | 105.97 | 9.99 |
| $Ni_{0.6}Co_{0.2}Fe_{0.2}$ | 112.61 | 102.93 | 9.68 |
| $Ni_{0.6}Co_{0.2}Mn_{0.2}$ | 107.81 | 101.11 | **6.71** |

# Supplemental Material for "Universal enhancement of vacancy diffusion by Mn inducing anomalous Friedel oscillation in concentrated solid-solution alloys"

## Method

First-principles density functional theory (DFT) calculations were performed based on the Projector Augmented Wave (PAW) method with the Perdew-Burke-Ernzerhof (PBE) exchange potential, as implemented in the Vienna Ab Initio Simulation Package (VASP) code [16,17]. The diffusion barriers and paths were investigated using the Climbing-Image Nudged Elastic Band (CI-NEB) method [18]. In this method, three intermediate images were used to optimize along the reaction path. The tetrahedron smearing method with Blöchl corrections [17] was used on fixed dimension/volume calculations to generate the density of states (DOS). The Brillouin zones were sampled using $K$ point grids with a uniform spacing of $2\pi \times 0.04$ A$^{-1}$. The model structures were fully optimized by using thresholds of $10^{-4}$ eV and 0.02 eV/Å for the total energy and force, respectively. The electron wavefunctions were expanded by the plane wave basis up to 400 eV. The preceding parameters were all carefully selected through pre-calculation to ensure that the results were accurate and that computational resource use was minimized as much as possible.

The structures were modeled utilizing the similar local approximation environment (SAE) method [14,15], which were generated by creating similar local atomic environments for all the lattice sites. The SAE method should be distinguished from the Special Quasirandom Structures (SQS) method, in which a cluster function is used to formulate the site occupation correlation function. Alternatively, the similarity function is presented to realize the correlation function in the SAE method. It has been shown that the optimal solution of the SAE method suffices as the optimal solution of the SQS method. The detail comparison between SAE and SQS has been discussed in our recent work [15]. In this study, the $Ni_{0.6}Co_{0.2}Co_{0.2}$, $Ni_{0.6}Co_{0.2}Fe_{0.2}$, and $Ni_{0.6}Co_{0.2}Mn_{0.2}$ CSAs are 2×3×1 supercells with a total atomic number of 120. **Fig.S1** shows the structure of CSAs supercells.

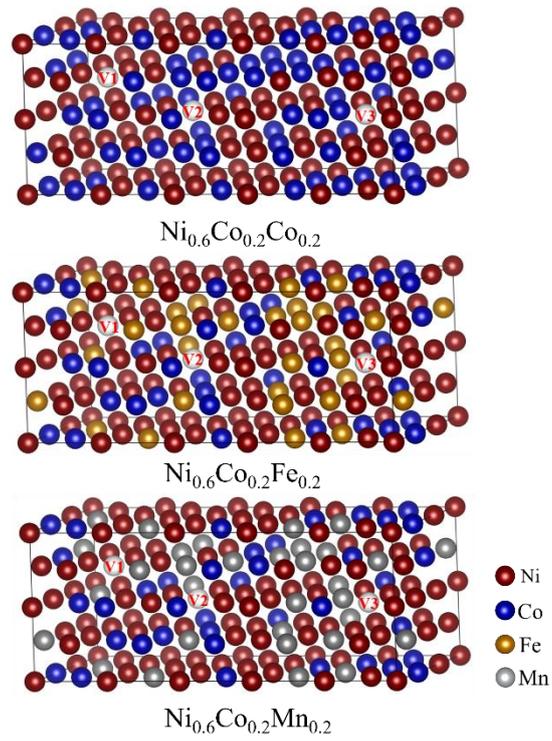

**Fig.S1.** Supercell structures of $Ni_{0.6}Co_{0.2}Co_{0.2}$, $Ni_{0.6}Co_{0.2}Fe_{0.2}$, and $Ni_{0.6}Co_{0.2}Mn_{0.2}$ by SAE method. Vacancies are randomly selected at the same position in the alloys.

# Diffusion Path

**Fig.S2** are the NEB calculation results of vacancy migration barriers based on different element types in 119-atom $Ni_{0.6}Co_{0.2}Co_{0.2}$, $Ni_{0.6}Co_{0.2}Fe_{0.2}$, and $Ni_{0.6}Co_{0.2}Mn_{0.2}$ supercells. In each alloy, all possible diffusion paths for three random single-vacancy are considered.

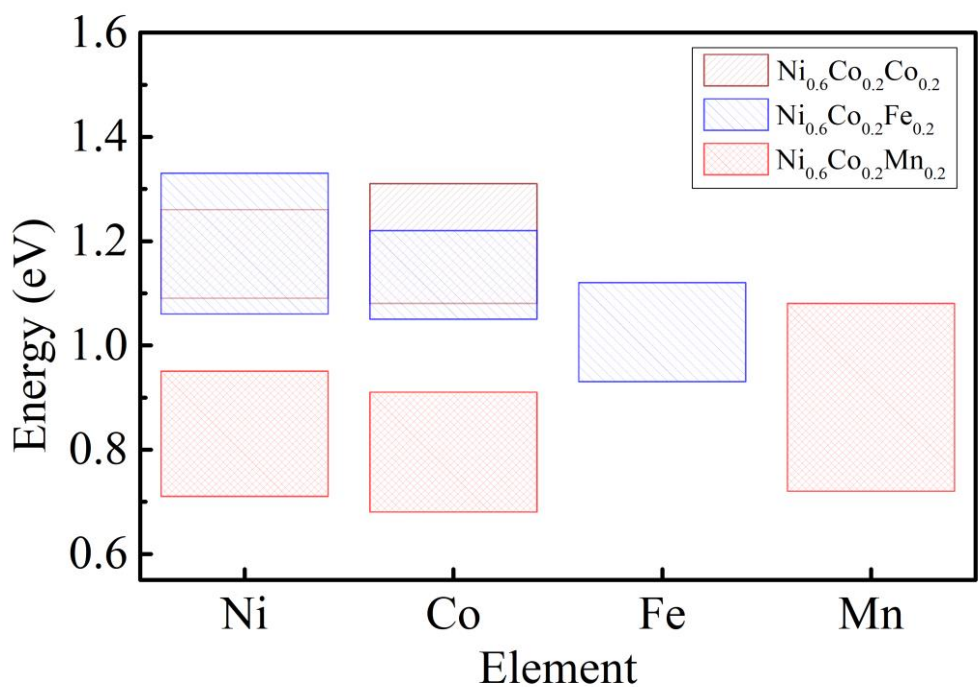

**FIG.S2.** The calculated migration barrier energy ranges of vacancies in $Ni_{0.6}Co_{0.2}Co_{0.2}$, $Ni_{0.6}Co_{0.2}Fe_{0.2}$, and $Ni_{0.6}Co_{0.2}Mn_{0.2}$ alloys. In each alloy, all possible diffusion paths for three random single-vacancy are summarized.

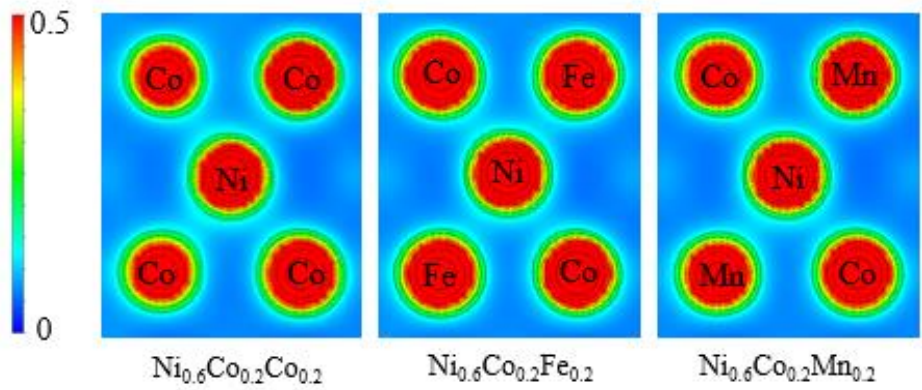

**Fig.S3.** Two-dimensional charge density distribution of the diffusing atom Ni and its 1nn atoms at the saddle point in $Ni_{0.6}Co_{0.2}Co_{0.2}$, $Ni_{0.6}Co_{0.2}Fe_{0.2}$, and $Ni_{0.6}Co_{0.2}Mn_{0.2}$, respectively.

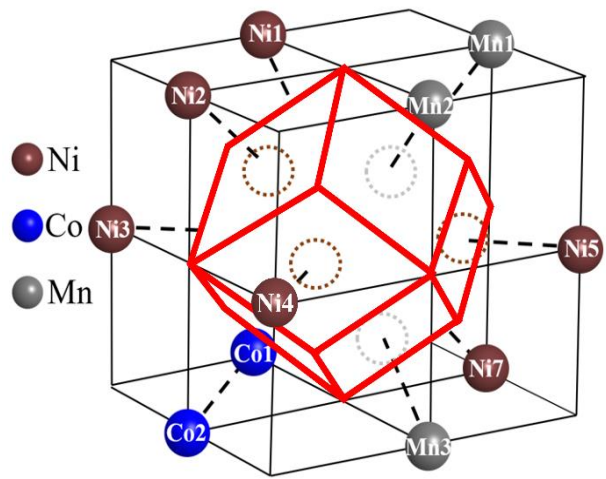

**Fig.S4.** The WS cell (shown by red lines) around the vacancy for the $Ni_{0.6}Co_{0.2}Mn_{0.2}$ supercell. The dash circles are the saddle points of different diffusion paths.